\documentclass[epjST]{svjour}
\usepackage[normalem]{ulem}
\def\slashchar#1{\setbox0=\hbox{$#1$}  
\dimen0=\wd0     
\setbox1=\hbox{/} \dimen1=\wd1  
\ifdim\dimen0>\dimen1   
\rlap{\hbox to \dimen0{\hfil/\hfil}} 
#1     
\else     
\rlap{\hbox to \dimen1{\hfil$#1$\hfil}} 
/      
\fi}
\newcommand{\dd}{\mathrm{d}}

\usepackage{amssymb}
\usepackage{amsmath}
\usepackage{comment}
\usepackage{multirow}
\usepackage{graphics}
\usepackage{color} 

\renewcommand\sout{\bgroup \color{blue}\ULdepth=-.5ex \ULset}

\begin{document}
\title{Speed of sound and quark confinement inside neutron stars}
\author{Micha\l{} Marczenko\inst{1}\fnmsep\thanks{\email{michal.marczenko@uwr.edu.pl}}}
\institute{Institute of Theoretical Physics, University of Wrocław,\\plac Maksa Borna 9, 50-204 Wrocław, Poland}
\abstract{Several observations of high-mass neutron stars (NSs), as well as the first historic detection of the binary neutron star merger GW17\-0817, have delivered stringent constraints on the equation of state (EoS) of cold and dense matter. Recent studies suggest that, in order to simultaneously accommodate a $2M_\odot$ NS and the upper limit on the compactness, the pressure has to swiftly increase with density and the corresponding speed of sound likely exceeds the conformal limit. In this work, we employ a unified description of hadron-quark matter, the hybrid quark-meson-nucleon (QMN) model, to investigate the EoS under NS conditions. We show that the dynamical confining mechanism of the model plays an important role in explaining the observed properties of NSs.
} 
\maketitle
\section{Introduction}\label{sec:introduction}

The equation of state (EoS) of strongly interacting matter is one of the key observables characterizing properties of matter under extreme conditions. It encodes information on the phase structure of Quantum Chromodynamics (QCD). One way of delineating the EoS is through the speed of sound, $c_s^2 \equiv \dd P /\dd \epsilon$. It has been conjectured that the speed of sound is bounded by $c_s^2 < 1/3$ (conformal bound). This conjecture was in fact confirmed in ab initio calculations of lattice QCD (LQCD) at finite temperature and small net-baryon density~\cite{Borsanyi:2012cr,Bazavov:2014pvz,Borsanyi:2013bia,Borsanyi:2010cj,Karsch:2006xs,Gavai:2004se}. However, the domain of low temperature and high net-baryon density is still inaccessible via the LQCD methods due to the infamous sign problem. This corner of the QCD phase diagram is of utmost importance for the understanding of the extraterrestrial observations, particularly for the study of neutron stars (NSs), their mergers~\cite{Bauswein:2018bma} and supernovae~\cite{Fischer:2017lag}. Until recently, the progress in constraining the EoS at low temperature and high density was driven mostly by several discoveries of high-mass NSs~\cite{Demorest:2010bx,Antoniadis:2013pzd,Fonseca:2016tux,Cromartie:2019kug}. These observations have set an important constraint on the maximum mass of a NS. More remarkably, the first ever detection of gravitational waves from the compact star merger GW170817~\cite{Abbott:2018exr}, the second detection from GW190425~\cite{Abbott:2020uma}, as well as the NICER observation of the millisecond pulsar PSR~J0030+451~\cite{Riley:2019yda,Miller:2019cac} delivered simultaneous measurement of masses and radii. Perturbative calculations of cold but dense QCD show that the speed of sound complies with the conformal bound~\cite{Kurkela:2009gj}, although they are reliable only at densities far beyond those realized in astrophysical objects. Contrary to these calculations, as well as the LQCD predictions, several recent analyses provided compelling reasons to expect that the conformal bound has to be violated at densities realized in the interior of NSs in order to support the observed NS properties~\cite{Bedaque:2014sqa,Tews:2018kmu}.

  The advancement of nuclear theory over the years has also tightened the constraints on the EoS over a wide range of densities. This has been achieved by systematic analyses of new astrophysical observations within simplistic approaches, such as the constant-speed-of-sound (CSS) model~\cite{Alford:2013aca} or multipolytropic class of EoSs~\cite{Hebeler:2013nza,Read:2008iy,Alvarez-Castillo:2017qki}. Recently, the interplay between the high-mass constraint (which requires high pressure) and the upper limit on the compactness from GW170817 event (which favors soft pressure) was used to derive a lower bound constraint on the maximal value of the speed of sound in the cold and dense matter EoS of a NS~\cite{Reed:2019ezm}. This new constraint strengthens previous expectations that the conformal bound is likely to be violated at densities realized inside NSs. 

  The lower bound of the speed of sound derived in~\cite{Reed:2019ezm} complies with constraints valid at various density regimes. This is particularly useful for determining a class of effective models in which the low-density and high-density regimes are not treated independently, but rather combined in a consistent unified framework. To this end, we employ the hybrid quark-meson-nucleon (QMN) model~\cite{Benic:2015pia,Marczenko:2017huu,Marczenko:2018jui,Marczenko:2019trv,Marczenko:2020jma} to quantify the EoS of cold and dense matter under NS conditions. The model has the characteristic feature that, at increasing baryon density, the chiral symmetry is restored within the hadronic phase by lifting the mass splitting between chiral partner states, before the quark deconfinement takes place. Quark degrees of freedom are included on top of hadrons, but their unphysical onset is prevented at low densities. This is achieved by an auxiliary scalar field which couples to both nucleons and quarks. This field serves as a momentum cutoff in the Fermi-Dirac distribution functions, thus it suppresses the unphysical thermal fluctuations of fermions, with the strength linked to the density. Our main focus is put on the role of the dynamical quark confinement in constraining the EoS of cold and dense matter under NS conditions. 

  This paper is organized as follows. In Sec.~\ref{sec:hybrid_qmn}, we introduce the hybrid quark-meson-nucleon model. In Sec.~\ref{sec:results}, we discuss the obtained numerical results on the equation of state under neutron-star conditions and neutron-star relations, and we confront them with recent observations and constraints. Finally, Sec.~\ref{sec:conclusions} is devoted to the summary and conclusions.

\section{Hybrid quark-meson-nucleon model}\label{sec:hybrid_qmn}

  In this section, we briefly introduce the hybrid QMN model for the chiral symmetry restoration and deconfinement phase transitions~\cite{Benic:2015pia,Marczenko:2017huu,Marczenko:2018jui,Marczenko:2019trv,Marczenko:2020jma}. The hybrid QMN model is composed of the baryonic parity doublet~\cite{Detar:1988kn,Jido:1999hd,Jido:2001nt} and mesons as in the Walecka model~\cite{Walecka:1974qa}, as well as quark degrees of freedom as in the standard linear sigma model~\cite{Scavenius:2000qd}. The spontaneous chiral symmetry breaking yields the mass splitting between the two baryonic parity partners, while it generates the entire mass of a constituent quark. In this work, we consider a system with $N_f=2$; hence, relevant for this study are the positive-parity nucleons, i.e., proton ($p_+$) and neutron ($n_+$), and their negative-parity partners, denoted as $p_-$ and $n_-$, as well as the up ($u$) and down ($d$) quarks. The fermionic degrees of freedom are coupled to the chiral fields $\left(\sigma, \boldsymbol\pi\right)$, the isosinglet vector-isoscalar field ($\omega_\mu$), and the vector-isovector field ($\boldsymbol \rho_\mu$). The important concept of statistical confinement is realized in the hybrid QMN model by introducing a medium-dependent modification of the particle distribution functions.

   \begin{table*}[t!]\begin{center}\begin{tabular}{|c|c|c|c|}
    \hline
    $\rho_0~$[fm$^{-3}$] & $E/A - m_+$ [MeV] & $K$~[MeV] & $E_{\rm sym}$~[MeV] \\ \hline\hline
    $0.16$               & $-16$             & 240       & 31 \\ \hline
    \end{tabular}\end{center}
    \caption{Properties of the nuclear ground state at $\mu_B = 923~$MeV and the symmetry energy used in this work.}
    \label{tab:external_params}
  \end{table*}

  The thermodynamic potential of the hybrid QMN model in the mean-field approximation reads~\cite{Marczenko:2020jma}
  \begin{equation}\label{eq:thermo_pot_iso}
   \Omega = \sum_{x=p_\pm,n_\pm,u,d}\Omega_x + V_\sigma + V_\omega + V_\rho + V_b \textrm.
  \end{equation}
  where the summation goes over the fermionic degrees of freedom. The spin degeneracy factor, $\gamma_x$ for nucleons is $\gamma_\pm=2$ for both positive- and negative-parity states, while the spin-color degeneracy factor for up and down quarks is $\gamma_q=2\times 3 = 6$. The kinetic part, $\Omega_x$, reads
  \begin{equation}\label{eq:thermokin}
   \Omega_x = \gamma_x \int\frac{\dd^3p}{\left(2\pi\right)^3} T \left[\ln\left(1-n_x\right) + \ln\left(1-\bar n_x\right)\right]\textrm,
  \end{equation}
  where the functions $n_x$ and $\bar n_x$ are the modified Fermi-Dirac distributions for nucleons
  \begin{subequations}\label{eq:cutoff_nuc}
  \begin{align}
         n_\pm &= \theta \left(\alpha^2 b^2 - \boldsymbol p^2\right) f_\pm \textrm,\\
    \bar n_\pm &= \theta \left(\alpha^2 b^2 - \boldsymbol p^2\right) \bar f_\pm
  \end{align}
  \end{subequations}
  and for quarks
  \begin{subequations}\label{eq:cutoff_quark}
  \begin{align}
        n_q &= \theta \left(\boldsymbol p^2-b^2\right) f_q \textrm,\\
   \bar n_q &= \theta \left(\boldsymbol p^2-b^2\right) \bar f_q \textrm,
  \end{align}
  \end{subequations}
  respectively. The model embeds the concept of statistical confinement through the modified Fermi-Dirac distribution functions, where $b$ is the expectation value of an auxiliary scalar field $b$ and $\alpha$ is a dimensionless model parameter. As demonstrated in Refs.~\cite{Benic:2015pia,Marczenko:2017huu,Marczenko:2018jui,Marczenko:2019trv,Marczenko:2020jma}, the parameter $\alpha$ plays also a crucial role in tuning the order of the chiral phase transition. From the definition of $n_\pm$ and $n_q$, it is evident that, in order to mimic the statistical confinement, the $b$ field should have a nontrivial vacuum expectation value, to suppress quark degrees of freedom at low densities in the confined and to allow for their population at high densities in deconfined phase. From Eqs.~\eqref{eq:cutoff_nuc} and~\eqref{eq:cutoff_quark}, one finds that the nucleons favor large $b$, whereas the quarks small $b$. The functions $f,\bar f$ are the standard Fermi-Dirac distribution functions for particle and antiparticle,
  \begin{subequations}
  \begin{align}
    f_x      &= \frac{1}{1+e^{\beta \left(E_x - \mu_x\right)}} \textrm,\\
    \bar f_x &= \frac{1}{1+e^{\beta \left(E_x + \mu_x\right)}}\textrm,
  \end{align}
  \end{subequations}
  respectively. $\beta$ is the inverse temperature, and the dispersion relation $E_x = \sqrt{\boldsymbol p^2 + m_x^2}$. The effective chemical potentials for $p_\pm$ and $n_\pm$ are defined as
  \begin{subequations}\label{eq:u_eff_had_iso}
  \begin{align}
    \mu_{p_\pm} &= \mu_B - g^N_\omega\omega - \frac{1}{2}g^N_\rho \rho + \mu_Q\textrm,\\
    \mu_{n_\pm} &= \mu_B - g^N_\omega\omega + \frac{1}{2}g^N_\rho \rho\textrm.
  \end{align}
  \end{subequations}
  The effective chemical potentials for up and down quarks are given by
  \begin{subequations}\label{eq:u_effq}
  \begin{align}
    \mu_u &= \frac{1}{3}\mu_B - g^q_\omega \omega - \frac{1}{2}g^q_\rho \rho + \frac{2}{3}\mu_Q\textrm,\\
    \mu_d &= \frac{1}{3}\mu_B - g^q_\omega \omega + \frac{1}{2}g^q_\rho \rho - \frac{1}{3}\mu_Q\textrm.
  \end{align}
  \end{subequations}
  In Eqs.~\eqref{eq:u_eff_had_iso}~and~\eqref{eq:u_effq}, $\mu_B$, $\mu_Q$ are the baryon and charge chemical potentials, respectively. 

  \begin{table*}[t!]\begin{center}\begin{tabular}{|c|c|c|c|c|c|}
    \hline
    $m_+~$[MeV] & $m_-~$[MeV] & $m_\pi~$[MeV] & $f_\pi~$[MeV] & $m_\omega~$[MeV] & $m_\rho~$[MeV] \\ \hline\hline
    939   & 1500  & 140     & 93      & 783        & 775 \\ \hline
    \end{tabular}\end{center}
    \caption{Physical vacuum inputs used in this work.}
    \label{tab:vacuum_params}
  \end{table*}

  The strength of $g^N_\omega$ is fixed by the nuclear saturation properties, while the value of $g^N_\rho$ can be fixed by fitting the value of symmetry energy~\cite{glendenning00:book}. The properties of the nuclear ground state and the symmetry energy are shown in Table~\ref{tab:external_params}. On the other hand, the nature of the repulsive interaction among quarks and their coupling to the $\omega$ and $\rho$ mean fields are still far from consensus. To account for the uncertainty in the theoretical predictions, one may treat the couplings $g^q_\omega$ and $g^q_\rho$ as free parameters. As demonstrated in Ref.~\cite{Marczenko:2020jma}, the repulsive quark-vector interaction has consequences for the phenomenological description of compact stellar objects of masses around $2M_\odot$. In the current work, however, we are interested in NSs with masses of $1.4M_\odot$. Thus, we neglect the repulsive quark-vector interactions and set $g_\omega^q = g_\rho^q = 0$ for simplicity of discussion.

  The effective masses of the chiral partners, $m_{p_\pm} = m_{n_\pm} \equiv m_\pm$, are given by
  \begin{equation}\label{eq:doublet_masses}
    m_\pm = \frac{1}{2} \left[ \sqrt{\left(g_1+g_2\right)^2\sigma^2+4m_0^2} \mp \left(g_1 - g_2\right)\sigma \right] \textrm.
  \end{equation}
  The positive-parity nucleons are identified as the positively charged and neutral $N(938)$ states, i.e., proton ($p_+$) and neutron ($n_+$). Their negative-parity counterparts, denoted as $p_-$ and $n_-$ are identified as $N(1535)$~\cite{Tanabashi:2018oca}. From Eq.~(\ref{eq:doublet_masses}), it is clear that the chiral symmetry breaking generates only the splitting between the two masses. When the chiral symmetry is restored, the masses become degenerate with a common finite mass $m_\pm\left(\sigma=0\right) = m_0$, which reflects the parity doubling structure of the \mbox{low-lying} baryons. Following the previous studies of the \mbox{parity-doublet-based} models~\cite{Benic:2015pia,Marczenko:2017huu,Marczenko:2018jui,Marczenko:2019trv,Marczenko:2020jma,Zschiesche:2006zj,Motornenko:2019arp,Mukherjee:2016nhb,Dexheimer:2012eu,Steinheimer:2011ea,Weyrich:2015hha,Sasaki:2010bp,Yamazaki:2019tuo,Mukherjee:2017jzi,Ishikawa:2018yey,Steinheimer:2010ib}, as well as recent lattice QCD results~\cite{Aarts:2017rrl,Aarts:2018glk}, we choose a rather large value, $m_0=700$~MeV. The couplings $g_1$ and $g_2$ in Eq.~\eqref{eq:doublet_masses} can be determined by fixing the fermion masses in the vacuum. Their values used in this work are summarized in Table~\ref{tab:vacuum_params}.

  The quark effective mass, $m_u = m_d \equiv m_q$, is linked to the sigma field as 
  \begin{equation}\label{eq:mass_quark}
    m_q = g_q \sigma \textrm.
  \end{equation}
  We note that in contrast to the baryonic parity partners (cf. Eq.~\eqref{eq:doublet_masses}), quarks become massless as the chiral symmetry gets restored. The value of the coupling $g_q$ in Eq.~\eqref{eq:mass_quark} can be determined by assuming the quark mass to be $m_q = 1/3~m_+$ in the vacuum.

  The potentials in Eq.~\eqref{eq:thermo_pot_iso} read
  \begin{subequations}\label{eq:potentials}
  \begin{align}
    V_\sigma &= -\frac{\lambda_2}{2}\left(\sigma^2 + \boldsymbol\pi^2\right) + \frac{\lambda_4}{4}\left(\sigma^2 + \boldsymbol\pi^2\right)^2 - \frac{\lambda_6}{6}\left(\sigma^2 + \boldsymbol\pi^2\right)^3- \epsilon\sigma \textrm,\label{eq:potentials_sigma}\\
    V_\omega &= -\frac{m_\omega^2 }{2}\omega_\mu\omega^\mu\textrm,\\
    V_\rho &= - \frac{m_\rho^2}{2}{\boldsymbol \rho}_\mu{\boldsymbol \rho}^\mu \textrm,\\
    V_b &= -\frac{\kappa_b^2}{2}b^2 + \frac{\lambda_4}{4}b^4 \textrm,\label{eq:potentials_b}
  \end{align}
  \end{subequations}
  where $\lambda_2 = \lambda_4f_\pi^2 - \lambda_6f_\pi^4 - m_\pi^2$, and $\epsilon = m_\pi^2 f_\pi$. $m_\pi$, $m_\omega$, and $m_\rho$ are the $\pi$, $\omega$, and $\rho$ meson masses, respectively, and $f_\pi$ is  the pion decay constant. The parameters $\lambda_4$ and $\lambda_6$ are fixed by the properties of the nuclear ground state. Numerical values of all model parameters are summarized in Table~\ref{tab:model_params}.

  \begin{table*}[t!]\begin{center}\begin{tabular}{|c|c|c|c|c|c|c|c|c|}
    \hline
    $\lambda_4$ & $\lambda_6f_\pi^2$ & $g^N_\omega$ &  $g^N_\rho$ & $g_1$ & $g_2$ & $g_q$ & $\kappa_b~$[MeV] & $\lambda_b$ \\ \hline\hline
     33.74          & 13.20           & 7.26       & 7.92 & 13.75 & 7.72 & 3.36& 155 & 0.074\\ \hline
    \end{tabular}\end{center}
    \caption{Numerical values of the model parameters. The values of $\lambda_4$, $\lambda_6$ and $g^N_\omega$ are fixed by the nuclear ground state properties, $g^N_\rho$ by the symmetry energy, and $g_q$ is fixed by the vacuum quark mass (see the text). The remaining parameters, $\kappa_b$ and $\lambda_b$ are fixed following Ref.~\cite{Marczenko:2017huu}.}
    \label{tab:model_params}
  \end{table*}

  In-medium profiles of the mean fields are obtained by extremizing the thermodynamic potential~in Eq.~\eqref{eq:thermo_pot_iso}. In the grand canonical ensemble, the~thermodynamic pressure is obtained from the thermodynamic potential as \mbox{$P = -\Omega + \Omega_0$}, where $\Omega_0$ is the value of the thermodynamic potential in the vacuum. The~net-baryon number density for a species $x$ is defined as
  \begin{equation}
    \rho^x_B = -\frac{\partial \Omega_x}{\partial \mu_B} \textrm,
  \end{equation}
  where $\Omega_x$ is the kinetic term in Eq.~\eqref{eq:thermokin}. The~total net-baryon number density~reads
  \begin{equation}
    \rho_B = \rho_B^{n_+} + \rho_B^{n_-} + \rho_B^{p_+} + \rho_B^{p_-} + \rho_B^{u} + \rho_B^{d} \textrm.
  \end{equation}

  In the following section the above hybrid QMN model equation of state of strongly interacting matter will be applied to identify properties of compact stellar objects such as NSs. The allowed range for the $\alpha$ parameter is $\alpha b_0 = 300 - 450~$MeV~\cite{Benic:2015pia,Marczenko:2017huu}, where $b_0$ denotes the vacuum expectation value of the $b$-field. Following our previous works, we choose four representative values within that interval: $\alpha b_0 = 350,~370,~400,~450~$MeV, to systematically study the phenomenology of compact stellar objects.

 \begin{figure}
    \resizebox{0.5\columnwidth}{!}{\includegraphics{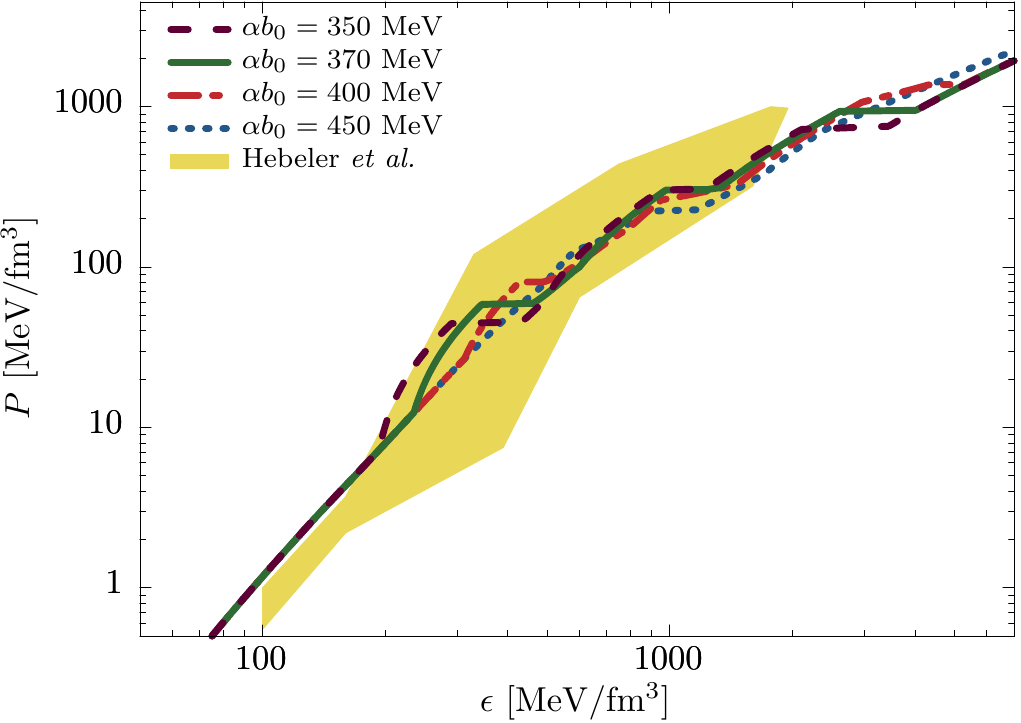}}
    \resizebox{0.496\columnwidth}{!}{\includegraphics{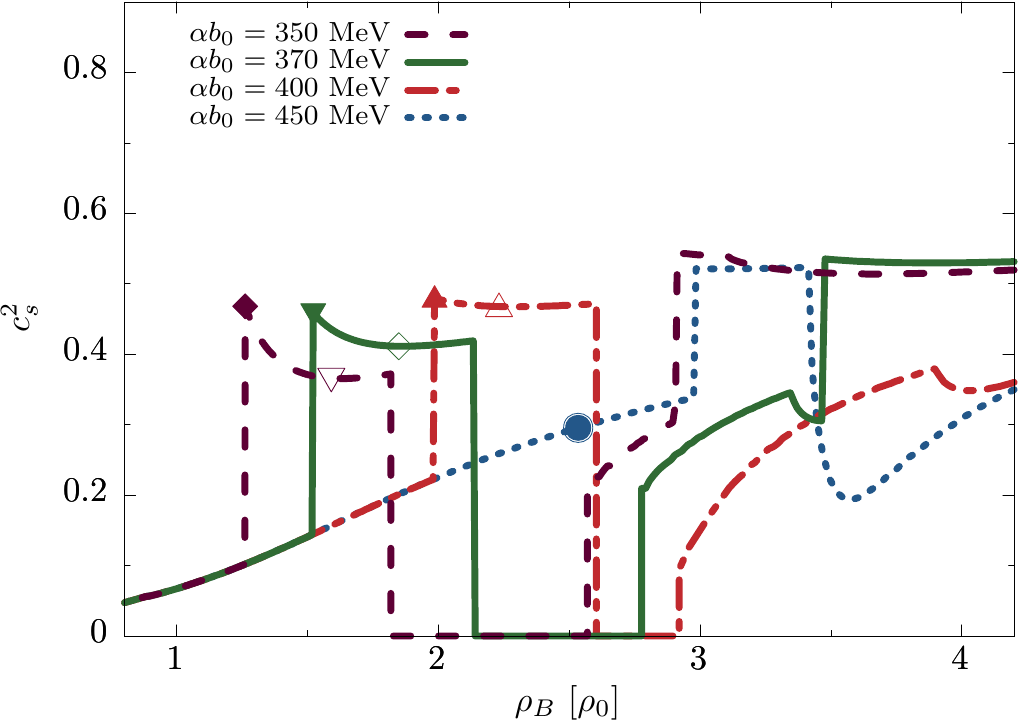}}
    \caption{Left panel: Thermodynamic pressure, $P$, under the NS conditions of $\beta$-equilibrium and charge neutrality, as a function of the energy density, $\epsilon$. Right panel: The square of the speed of sound, $c_s^2$, as a function of the baryon density, $\rho_B$, in the units of saturation density, in the vicinity of the chiral phase transition. The first-order phase transitions are seen in the left panel as plateaux of constant pressure and as vanishing speed of sound in the right panel. In the left panel, the yellow-shaded region marks the constraint obtained by Hebeler {\it et al.}~\cite{Hebeler:2013nza}. In the right panel, the open symbols show the central densities of corresponding $1.4M_\odot$ NSs, and the filled symbols show the corresponding densities at which the maximal value of the speed of sound within $1.4M_\odot$ NSs is reached. Results in both panels are obtained for $m_0 = 700~$MeV and four representative values of the parameter $\alpha$.}\label{fig:p_e_cs2}
  \end{figure}

\section{Results}\label{sec:results}

  The composition of NS matter requires $\beta$-equilibrium, as well as the charge neutrality condition. To this end, we include electrons and muons as gases of free relativistic particles. In the left panel of Fig.~\ref{fig:p_e_cs2}, we show the calculated EoSs for $m_0=700~$MeV, as functions of the energy density, for different values of the $\alpha$ parameter, namely $\alpha b_0=350~$MeV (dashed, magenta line), $\alpha b_0=370~$MeV (solid, green line), $\alpha b_0=400~$MeV (dash-dotted, red line), and $\alpha b_0=450~$MeV (dotted, blue line). Shown EoSs feature chiral phase transitions, defined as a jump in the $\sigma$-field expectation value, which causes the parity partners to become almost degenerate with mass $m_\pm = m_0$. The mechanism of statistical confinement introduced in the previous section has a prominent impact on the stiffness of class of EoSs obtained in the model. Namely, higher values of the parameter $\alpha$ yield weaker first-order transitions, triggered at higher densities, which eventually becomes a smooth crossover. As the density increases, the EoSs feature another two sequential transitions. First, associated with the onset of the down quark, and second, associated with the onset of up quark, after which the matter is fully deconfined and comprised solely of quarks. We note that the sequential appearance stays in contrast to the isospin-symmetric case, where quarks deconfine simultaneously, owing to the isospin symmetry~\cite{Marczenko:2020jma}. Such separation of the chirally broken and the deconfined phase might indicate the existence of a quarkyonic phase, where the quarks are partly confined to form a Fermi sphere, but the relevant degrees of freedom around the Fermi surface remain the nucleons with the restored chiral symmetry~\cite{Hidaka:2008yy,McLerran:2008ua,Andronic:2009gj,McLerran:2018hbz,Jeong:2019lhv,Zhao:2020dvu}.

  We note that the discussed EoSs are in good agreement with the maximum-mass constraint obtained by using a multi-polytrope ansatz for the EoS above the saturation density, shown as yellow-shaded region in the left panel of Fig.~\ref{fig:p_e_cs2}. Interestingly, the chiral phase transitions and the deconfinement of down quark lie within the region. We note that, in principle, the inclusion of the quark-vector repulsive coupling has an impact only on the high density part of the EoS, when compared to the case with vanishing coupling. Namely, it extends the hadronic branch to higher densities, and simultaneously, shifts the appearance of the quarks (see~\cite{Marczenko:2020jma} for details). The values of the density jumps associated with the chiral symmetry restoration, and consequent onset of down and up quarks featured in the class of EoS obtained in the model are shown in Table.~\ref{tab:jumps}.

    \begin{table*}[t!]\begin{center}\begin{tabular}{|c|c|c|c|}
    \hline
    \multicolumn{4}{|c|}{$\alpha b_0~$ [MeV]}         \\ \hline
    $350$                    & $370$           & $400$           & $450$    \\ \hline\hline
    $1.82 - 2.60$   & $2.14 - 2.76$   & $2.61 - 2.92$   & $3.56$ \\
    $4.98 - 6.11$   & $5.84 - 6.21$   & $5.10$          & $4.82 - 6.04$ \\
    $9.21 - 13.58$  & $11.03 - 15.42$ & $16.40 - 19.25$ & $10.84$ \\ \hline
    \end{tabular}\end{center}
    \caption{Baryon density ranges of the coexistence phases associated with the chiral restoration (top), onset of down (middle) and up (bottom) quark under the neutron-star conditions, in terms of saturation density units, $\rho_0$, for $m_0=700~$MeV and different values of $\alpha b_0$. In the cases were transitions proceed as smooth crossovers, a single value is given.}
    \label{tab:jumps}
  \end{table*}

  In the right panel of Fig.~\ref{fig:p_e_cs2}, we show the calculated speeds of sound, $c_s^2 \equiv \dd P / \dd \epsilon$, in the units of the speed of light, as a function of the baryon number density, $\rho_B$. For $\alpha b_0=350,~370,~400~$MeV, the coexistence phases of chirally broken and restored phases due to first order phase transitions are seen as regions of vanishing speed of sound. For $\alpha b_0=450~$MeV, the chiral phase transition proceeds as a smooth crossover and is seen as a dip around $3.5\rho_0$. We note that the sequential appearance of down and up quarks are triggered at higher densities and are not shown in the figure. The notable rapid increase of the speed of sound in each curve, before the chiral phase transition takes place, is a result of the stiffening mechanism that arises due to the statistical confinement implemented in the model (cf. Eq.~\eqref{eq:cutoff_nuc}).

\begin{figure}
    \resizebox{0.5\columnwidth}{!}{\includegraphics{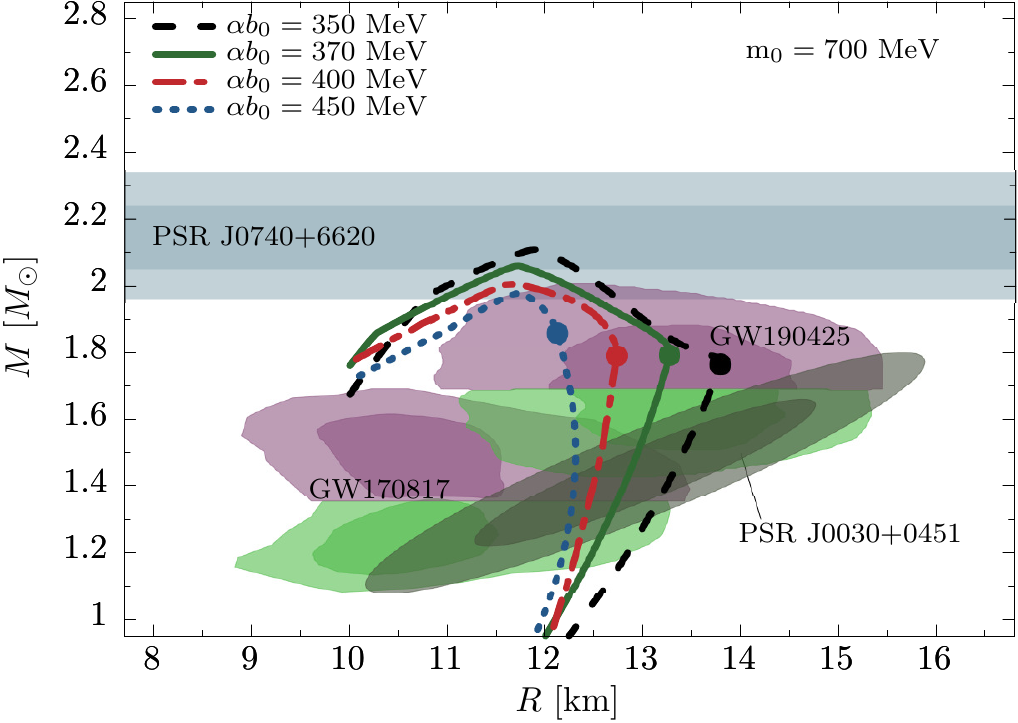}}
    \resizebox{0.496\columnwidth}{!}{\includegraphics{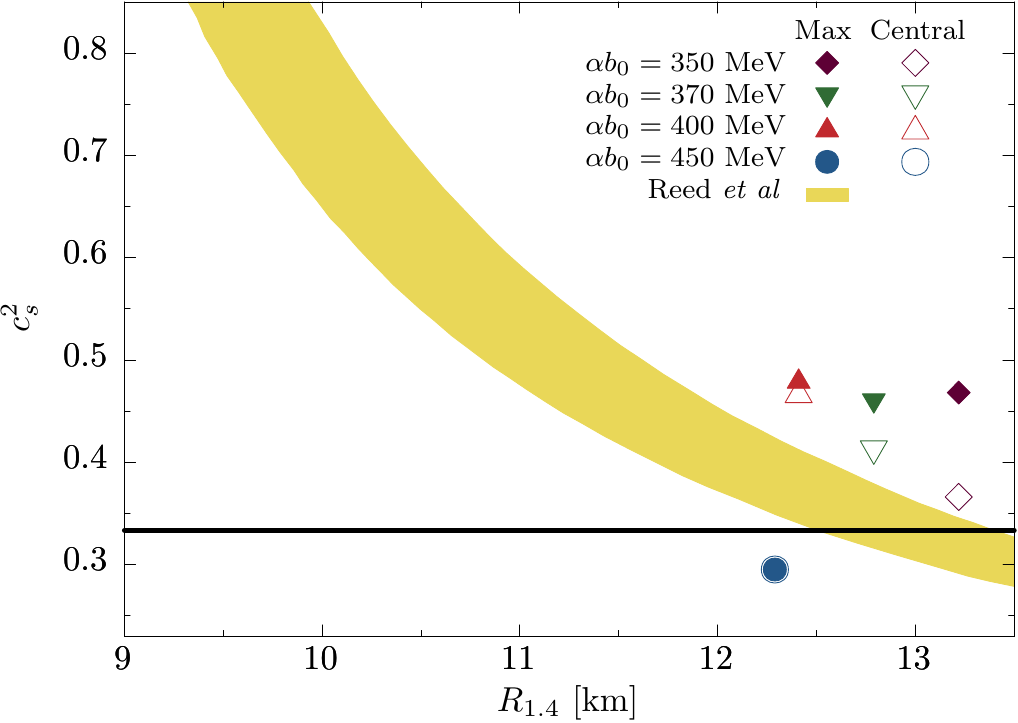}}
    \caption{Left panel: Mass-radius sequences obtained for compact star solutions of the TOV equations. The inner (outer) gray band shows the 68.3\%~(95.4\%) credibility regions for the mass of PSR J0740+6620~\cite{Cromartie:2019kug}. The inner (outer) green and purple bands show 50\%~(90\%) credibility regions obtained from the recent GW170817~\cite{Abbott:2018exr} and GW190425~\cite{Abbott:2020uma} events for the low- and high-mass posteriors. The inner (outer) black region corresponds to the mass and radius constraint at 68.2\% (95.4\%) obtained for PSR J0030+0451 by the group analyzing NICER X-ray data~\cite{Miller:2019cac}. The circles mark the coexistence of the chirally broken and chirally restored phases. Right panel: The maximal value of the square of the speed of sound, $c_s^2$, within a $1.4M_\odot$ NS, as a function of its radius, $R$. The central values of the speed of sound (open symbols) and maximal values within $1.4M_\odot$ and corresponding radii are shown. The yellow-shaded region represents the lower bound for the maximum value of the speed of sound inside a $1.4M_\odot$ neutron star derived in Ref.~\cite{Reed:2019ezm}. The black horizontal line shows the conformal value $c_s^2=1/3$. Results in both panels are obtained for $m_0 = 700~$MeV, and four representative values of the parameter $\alpha$.}\label{fig:m_r_cs2}
  \end{figure}

  We use the obtained EoSs in the mean-field approximation to solve the general-relativistic Tolman-Oppenheimer-Volkoff (TOV) equations for spherically symmetric objects at zero temperature, in order to obtain the mass-radius relations for NSs. In the left panel of Fig.~\ref{fig:m_r_cs2}, we show the mass-radius relations calculated from the introduced EoSs, together with the state-of-the-art constraints: the high-mass measurement of the PSR J0740+6620~\cite{Cromartie:2019kug}, two recent GW170817~\cite{Abbott:2018exr} and GW190425~\cite{Abbott:2020uma} events, and the mass-radius constraint obtained for J0030+0451~\cite{Miller:2019cac}. The agreement with all of the constraints is good. Notably, the chiral phase transitions (shown as circles) are featured within the mass and radius region accessible by GW190425, at around $1.8M_\odot$. We note that the presented results are calculated for vanishing quark-vector interactions, i.e., $\chi=0$. Finite value of the parameter $\chi$, leads to overall improvement of the phenomenological description of compact objects around $2M_\odot$ (see Ref.~\cite{Marczenko:2020jma} for details).

  In general, there is a tension between the constraints from high-mass measurements and gravitational-wave observations. On the one hand, it is long known that the existence of $2M_\odot$ NSs is supported by EoSs that are stiff enough at high densities above $2\rho_0$. On the other hand, the deformability constraint favors EoSs that are soft around $\sim1-2\rho_0$. In~\cite{Reed:2019ezm}, the interplay between constraints from these two types of measurements was used to derive a lower limit on the speed of sound in a $1.4M_\odot$ NS within a class of simplistic constant-speed-of-sound (CSS) EoSs. The constraint is shown in the right panel of Fig.~\ref{fig:m_r_cs2} as a function of radius, $R_{1.4}$, of $1.4M_\odot$ NS (yellow-shaded region). The speed of sound monotonically decreases as $R_{1.4}$ increases. However, in principle, it is rather unlikely that the speed of sound is independent of density. Thus, if at some density the speed of sound is below the value needed to comply with the maximum-mass constraint, then it may have to be larger than the desired value of the constraint, at other densities. Therefore, the constraint places a lower estimate for the maximum of the speed of sound in dense NS matter. We note that the constraint also implies that the maximal value of the speed of sound has to exceed the conformal bound, i.e., $c_s^2=1/3$ (horizontal, black line).

  In the figure, we also show the maximal values of $c_s^2$ (filled symbols) within $1.4M_\odot$ obtained for each parametrization in the hybrid QMN model, together with corresponding central values (open symbols). For $\alpha b_0=350,~370,~400~$MeV, the values are not only above the conformal limit, but they also lie above the constraint. Notably, the maximal values of the speed of sound are obtained at densities where the stiffening of the EoS set in. This is seen in the right panel of Fig.~\ref{fig:p_e_cs2}, where the maximal values of $c_s^2$ in $1.4M_\odot$ are shown as filled symbols. On the other hand, the maximal value of the speed of sound for $\alpha b_0=450~$MeV does not exceed the conformal value. In this case, $c_s^2$ rises monotonically even beyond the central density of the $1.4M_\odot$ NS, and the stiffening sets in at higher densities (see right panel of Fig.~\ref{fig:p_e_cs2}). The obtained radii, central baryon densities, and the maximal speeds of sound of $1.4M_\odot$ neutron stars for all parametrizations are provided in Table~\ref{tab:r_lambda_cs2}. Seemingly, sufficient stiffening of the EoS at densities just above the saturation density, realizable in $1.4M_\odot$ NS, is a trademark that is required in order to comply with the constraint from Ref.~\cite{Reed:2019ezm}. In the hybrid QMN model, it is provided through the dynamical mechanism of confinement which strength in linked to the density. We note that this effect is also featured in other effective approaches, such as the relativistic density-functional models with excluded nucleon volume~\cite{Typel:2016srf} and the class of CSS hybrid EoSs~\cite{Alford:2017qgh}. However, too extreme stiffening can become at certain tension with the analysis of GW170817~\cite{Abbott:2018exr}. In principle, this tension could be resolved, e.g., by a strong phase transition occurring in the mass range relevant for GW170817~\cite{Paschalidis:2017qmb}. We note that because in the hybrid QMN model the chiral phase transition featured around $1.8M_\odot$, the obtained $1.4M_\odot$ NSs are composed solely of nuclear matter with broken chiral symmetry. This is seen in the right panel of Fig.~\ref{fig:p_e_cs2}, where the central values of the speed of sound (open symbols) of $1.4M_\odot$ NSs are obtained at baryon densities below the chiral phase transition region (plateaux of vanishing speed of sound). Therefore, the inclusion of the statistical confinement has important implications already at densities before the quarks deconfine. This is even more pronounced in a class of models with sequential chiral and deconfinement phase transitions, such as the hybrid QMN model. This may have important phenomenological implications for the study of multi-messenger astronomy and heavy ion collisions (HIC)~\cite{Sasaki:2019jyh}.

  \begin{table}[t!]\begin{center}\begin{tabular}{|c||c|c|c|}\hline
       $\alpha b_0~$[MeV] & $R_{1.4}$~[km] & $\rho_B~[\rho_0]$   & $c_s^2$ \\ \hline\hline
       350                & 13.22          & 1.59                & 0.468   \\ \hline
       370                & 12.79          & 1.85                & 0.459   \\ \hline
       400                & 12.41          & 2.23                & 0.480   \\ \hline
       450                & 12.29          & 2.53                & 0.295   \\ \hline
      \end{tabular}\end{center}
      \caption{Values of radius, central baryon density, and maximal value of the speed of sound of $1.4M_\odot$ NSs obtained in the hybrid QMN model for different values of $\alpha$ parameter for $m_0=700~$MeV.}
      \label{tab:r_lambda_cs2}
  \end{table}

\section{Conclusions}\label{sec:conclusions}
  
  The progress in constraining the equation of state (EoS) of cold and dense matter under extreme conditions is driven mostly through modern multi-messenger-astronomy observations. Several high-mass neutron star (NS) observations~\cite{Demorest:2010bx,Antoniadis:2013pzd,Fonseca:2016tux,Cromartie:2019kug} and the first ever detection of the gravitational waves from the compact-star merger GW170817~\cite{Abbott:2018exr} have delivered powerful constraints on the NS mass-radius profile independently. In fact, there is an apparent tension between the high-mass constraint (which requires high pressure) and the upper limit on the compactness from GW170817 event (which favors soft pressure). Recently, the interplay between them was used to derive a lower bound constraint on the maximal value of the speed of sound in the cold and dense matter EoS of a NS~\cite{Reed:2019ezm}. This new constraint strengthens previous expectations that the conformal bound is likely to be violated at densities realized inside NSs~\cite{Bedaque:2014sqa,Tews:2018kmu}. Because such constraint characterizes different density regimes, it is of particular use for a class of effective models in which the low-density and high-density regimes are not treated independently, but rather combined in a consistent unified framework, such as, e.g.,~\cite{Marczenko:2020jma,Bastian:2015avq,Bastian:2018wfl}.

  In this study, we have utilized the hybrid quark-meson-nucleon (QMN) model to quantify the equation of state (EoS) of cold and dense matter. The model unifies the thermodynamics of quark and hadronic degrees of freedom. The interplay between the quark confinement and the chiral symmetry breaking is embedded in a dynamical way into a single unified framework. Within this approach, we have systematically investigated the EoS of cold and dense asymmetric matter under NS conditions. We have constructed the mass-radius relations based on solutions of the Tolman-Oppenheimer-Volkoff (TOV) equations.
  
  We have shown that in order to comply with modern constraints from multi-messenger astronomy, a rapid increase of pressure is required at densities inside a $1.4M_\odot$ NS. In the hybrid QMN model, such stiffening is naturally connected to the dynamical mechanism of confinement which strength in linked to the density. This result highlights the fact that the confinement plays a crucial role in the phenomenology of matter under extreme conditions, even at densities smaller than the density at which the system undergoes a hadron-to-quark phase transition. We note that in Ref.~\cite{Marczenko:2017huu}, it was shown that the confinement mechanism of the hybrid QMN model leads to strengthening of the chiral phase transition when compared to pure quark or hadronic models without confinement. This motivates further study and possible applications of the model, not only in the context of multi-messenger astronomy, but also in the context of heavy ion collisions (HIC) at finite temperature and density, where a novel signature of chiral symmetry restoration within the dense hadronic sector in dilepton production was recently proposed~\cite{Sasaki:2019jyh}.

\begin{acknowledgement}
  I acknowledge fruitful discussions and helpful comments from Krzysztof Redlich and Chihiro Sasaki. This work was partly supported by the Polish National Science Center (NCN), under Preludium Grant No. UMO-2017/27/N/ST2/01973.
\end{acknowledgement}

\end{document}